\documentclass[12pt]{article}

\usepackage{natbib,upgreek}
\usepackage{multirow,colonequals}
\usepackage{bm,amsmath,amsfonts,amssymb,graphicx}

\newcommand{\EX}{\mathbb{E}}

\newcommand{\bTh}{\boldsymbol{\Theta}}

\begin{document}
\title{Detecting British Columbia Coastal Rainfall Patterns by Clustering Gaussian Processes}

\author{Forrest Paton and Paul D. McNicholas}

\date{{\small Department of Mathematics and Statistics, McMaster University, Hamilton, Ontario, Canada.}}
\maketitle

\begin{abstract}

Functional data analysis is a statistical framework where data are assumed to follow some functional form. This method of analysis is commonly applied to time series data, where time, measured continuously or in discrete intervals, serves as the location for a function's value. Gaussian processes are a generalization of the multivariate normal distribution to function space and, in this paper, they are used to shed light on coastal rainfall patterns in British Columbia (BC). Specifically, this work addressed the question over how one should carry out an exploratory cluster analysis for the BC, or any similar, coastal rainfall data. An approach is developed for clustering multiple processes observed on a comparable interval,  based on how similar their underlying covariance kernel is. This approach provides interesting insights into the BC data, and these insights can be framed in terms of El~Ni\~{n}o and La~Ni\~{n}a; however, the result is not simply one cluster representing El~Ni\~{n}o years and another for La~Ni\~{n}a years. From one perspective, the results show that clustering annual rainfall can potentially be used to identify extreme weather patterns.

\noindent \textbf{Keywords}:
British Columbia; Clustering; Coastal Rainfall; El~Ni\~{n}o; Extreme weather; Gaussian processes; La~Ni\~{n}a; Mixture model.
\end{abstract}

\section{Introduction}
In contrast to predictable yearly seasonal changes, El~Ni\~{n}o, a well studied teleconnection, does not occur at regular intervals. This is of particular interest for policy makers, businesses, and people who rely on calculable, foreseeable weather patterns. For example, seasonal changes can specifically alter food production plans, such as when to deploy fishing vessels or harvest crops. 
While El~Ni\~{n}o is primarily categorized through warming temperatures in the eastern and equatorial Pacific Ocean, its effects can be seen around the globe through teleconnections. Teleconnections are an environmental phenomena that describe correlated large-scale atmospheric changes over non-contiguous geographic regions. While some teleconnections are well established, others rely on observing statistical irregularities \citep{gudmundson2003,ward}; namely, El~Ni\~{n}o's effect on precipitation patterns. Sir Gilbert Walker, a 20th century English scientist, for example, first identified the link between Asian monsoons and Pacific coastal barometer readings \citep{gudmundson2003}. However, the study of teleconnections, such as precipitation patterns, is notoriously complex because of intricate spatial and temporal correlations. While spatial data is often modelled under the assumption that geographical points near one another share more information than those far apart, work has been done to model both nearby and remote geographical correlations. \cite{hewitt2018} predict precipitation in Colorado by modelling both locally available data and remote Pacific Ocean sea surface temperatures. Understanding El~Ni\~{n}o's effect on distant precipitation can shed light on these patterns. Specifically, classifying irregular precipitation patterns in {the Americas} can help understand El~Ni\~{n}o's impact on local weather systems. 

A question of particular interest is: which years exhibit distinct rainfall patterns? %
For prediction, knowing how Pacific Ocean changes affect rainfall can have significant implications \citep{ogorman15}. A model for clustering yearly rainfall data will also give insight to research on the mechanistic properties of teleconnections. This has overlap in the statistical field of functional data clustering, where data are assumed to follow some functional form. Herein, mixture model-based clustering provides an effective approach to cluster data, and Gaussian Processes provide a model for rainfall. Much recent work has been done modelling climatological data with spatial and time dependence \citep{ armal18, cabral19, gupta16}. Specifically, the use of Gaussian processes (GPs) gives a probabilistic starting point. A GP is a stochastic process that generalizes a finite-dimensional normal distribution to function space. GPs have been used to successfully solve complex non-linear regression and classification problems \citep{rass, roberts2013}. When multiple functions exist on the same interval, usually compact $[0,T]$ and finite, it can be useful to classify them into a finite number of mutually exclusive groups. Here the process would be defined on the index set time, specifically the months of the year.

\section{Background}
\subsection{Gaussian Process}
\cite{rass} defines a GP as: ``\ldots a collection of random variables, any finite number of which have a joint Gaussian distribution". A GP $\mathbf{f}(\mathbf{x})$ creates a set of random variables evaluated at $\mathbf{x}$. In essence, a  GP is a distribution over functions, i.e.,
\begin{equation}
\mathbf{f}(\mathbf{x}) \sim \text{GP} \big( m(\mathbf{x}), \ k(\mathbf{x}, \mathbf{x}^{\top}) \big),
\end{equation}
where
$m(\mathbf{x}) = \EX[\mathbf{f}(\mathbf{x})]$ and
$k(\mathbf{x}, \mathbf{x}^{\top}) = \EX[(\mathbf{f}(\mathbf{x}) - m(\mathbf{x}))(\mathbf{f}(\mathbf{x}^{\top}) - m(\mathbf{x}^{\top}))]$
are the mean function and covariance kernel, respectively, $\mathbf{x}=({x}_1,\ldots,{x}_n)^{\top}$ is the function's index, and $\mathbf{f}(\mathbf{x})=(f_1, \ldots, f_n)^{\top}$ is the function's output, i.e., $(x_i, f_i)$ is a point in $\mathbb{R}^2$. A common GP, and the kernel considered in this paper, is defined with mean function $\textbf{0}$ and squared exponential (SE) covariance function
\begin{equation}\label{se}
    k(x_i, x_j) =  \sigma^2\exp\left\{-\frac{1}{2l^2} (x_i - x_j)^2\right\},
\end{equation}
where $\sigma^2$ and $l$ are hyper-parameters that control the shape of the process; specifically, $\sigma^2$ controls the amount of variation in $\mathbf{f}(\mathbf{x})$ and $l$, the length-scale parameter, controls the correlation.
 The SE kernel is a widely used \citep{rass}. One convenient property of the SE kernel is infinite differentiability, which is useful because the first derivative is needed for hyper-parameter estimation. Here, the use of the term hyper-parameter refers to a set of parameters that make up the ``non-parametric model". That is, the hyper-parameters only appear in the model's prior and, as shown later in \eqref{gplike}, are integrated out of the final model (posterior). 

Although the SE is the most popular, different kernel functions can be used. Because the kernel is needed to populate a multivariate normal distribution covariance matrix, the kernel is restricted to produce positive semi-definite matrices. While the SE covariance kernel is a popular choice, modelling the shape of the covariance kernel is an open ended problem. One systematic solution can be formed by considering a Bayesian model selection framework as discussed in \cite{rass}. \cite{duvenaud2013} also provide a solution by considering automatic kernel selection through searching over appropriate kernel structures. For the SE covariance kernel, $\sigma^2$ controls the height or the amplitude of the GP and $l$ controls the correlation between observations. 
This kernel is used to construct a  matrix, $\textbf{K}$, which will serve as the covariance matrix in a multivariate normal distribution introduced in the next section:
\begin{align*}
\textbf{K} = \begin{pmatrix} k(x_1, \ x_1) & \cdots & k(x_1, \ x_n) \\ \vdots  &  \ddots & \vdots\\
k(x_n, \ x_1) & \cdots & k(x_n, \ x_n)
\end{pmatrix}.
\end{align*}

While a GP is defined on the entire real line, we only observe a finite number $n$ of realizations $\mathbf{x}=({x}_1,\ldots,{x}_n)^{\top}$, and corresponding $\mathbf{y}=({y}_1,\ldots,{y}_n)^{\top}$, where $\mathbf{y} \triangleq \mathbf{f}(\mathbf{x})$. The vector $\mathbf{x}$ is commonly called the input and represents the location of the the process, i.e., observation $y_i = f(x_i)$. The vector $\mathbf{y}$ is referred to as the output, and is the function evaluated at location $\mathbf{x}$. This allows for a generalization to a multivariate normal distribution via $\mathbf{f}(\mathbf{x}) \sim \mathcal{N} \big( \textbf{0}, \ \textbf{K}\big)
$. This is possible because marginalizing a Gaussian distribution is trivial: the resulting distribution is Gaussian and we can ignore the ($x, y$) pairs that are unobserved or missing. Here changing the kernel affects the shape of the function, effectively controlling the magnitude of which observations $x_i$ and $x_j$ are correlated. Formulating the problem in this way, we can see the kernel is our prior on the function space, and the (marginal) likelihood for the GP comes after conditioning on the realized points.
As shown in the next section, the hyper-parameters are often estimated to maximize the likelihood of the GP.

\subsection{Likelihood}
The previous section introduced the kernel function and how it relates to a prior on function space and how the hyper-parameters affect the correlation between the input $\mathbf{x}$ and outcome $\mathbf{y}$. Now, the likelihood for a GP will be introduced and strategies for choosing the hyper-parameters will be illustrated.
From the definition of a GP, $\mathbf{y} \sim \mathcal{N}(\mathbf{0}, \textbf{K})$, which will be shown formally below.
First, let $\mathbf{f} \sim \mathcal{N}(\mathbf{0}, \textbf{K})$, where $\textbf{K}$ is the covariance matrix constructed from the SE kernel shown in \eqref{se}.

The likelihood for a GP is conditioned on the observed values to obtain a marginal likelihood, using $\phi$ to denote the normal density function:
$$
p(\mathbf{f} \ |\ \mathbf{x}) = \underbrace{\phi(\mathbf{f} \ |\ \mathbf{0}, \textbf{K})}_\text{function prior},\qquad\qquad
p(\mathbf{y}\ |\ \mathbf{f}) = \underbrace{\prod_{i=1}^n\phi(y_i\ |\ f_i)}_\text{likelihood}.
$$
We get the marginal for $\mathbf{y}$ by using Bayes' rule and integrating over $\mathbf{f}$:
\begin{align}
    p(\mathbf{y} | \mathbf{x}) &= \int p(\mathbf{y} | \mathbf{f}, \mathbf{x}) \ p(\mathbf{f} \ | \mathbf{x})d\mathbf{f}.\label{gplike}
\end{align}
Taking the natural logarithm  of \eqref{gplike} gives 
\begin{align}\label{likelihood_gp}
\log p(\mathbf{y} | \mathbf{x}) &= -\frac{1}{2} \Big\{\mathbf{y}\textbf{K}^{-1}\mathbf{y}^{\top} + \log|\textbf{K}| + n\log2\pi\Big\}.
\end{align}
This likelihood can be broken down into three main components, the data fit term, model complexity term, and a constant term:
\begin{align}
   \log p(\mathbf{y} | \mathbf{x}) &=-\frac{1}{2} \Big\{\underbrace{\mathbf{y}\textbf{K}^{-1}\mathbf{y}^{\top}}_\text{data fit} \ + \underbrace{\log |\textbf{K}|}_\text{complexity}+
    \underbrace{n\log2\pi}_\text{constant} \Big\}.
\end{align}
The data fit and complexity component share an interesting tradeoff. For small length-scale values $l$, the model will fit the data well and the data fit component will be small. However, points will not be considered ``near" each other, resulting in a high model complexity. Conversely, if $l$ is large (suggesting no correlation between points), then the the complexity will be small but the data fit term will be large \citep{murphy2012}. This is because the SE kernel will converge to $\sigma^2$, turning $\textbf{K}$ into a diagonal matrix. Because GPs have these inherent penalty terms for over- and under-fitting, cross validation is generally not used to estimate kernel hyper-parameters. 

\subsection{Predictive Distribution}

GPs are commonly used in supervised regression tasks for their ability to non-parametrically approximate complex functions and solve functional engineering problems \citep{ecepaper}. 
It is often of interest to infer the function's value outside of the paired training data ($\mathbf{x}, \mathbf{y}$). To do this, a predictive distribution can be constructed. Let $\mathbf{y}_*=\mathbf{f}(\mathbf{x}_*)$ be the unobserved outputs to be inferred at locations $\mathbf{x}_*$. The joint distribution can be derived through probabilistic terms, 
\begin{equation}
     \mathbf{f}(\mathbf{x}) \sim \mathcal{N} \left( \mathbf{0}, \\  \textbf{K}(\mathbf{x}, \mathbf{x})   \right),
\end{equation}
\begin{equation}
     \mathbf{f}(\mathbf{x}_*) \sim \mathcal{N} \left( \mathbf{0}, \\  \textbf{K}(\mathbf{x}_*, \mathbf{x}_*)   \right),
\end{equation}
\begin{equation}\label{joint_g}
    \begin{bmatrix} \mathbf{f}(\mathbf{x}) \\ \mathbf{f}(\mathbf{x}_*) \end{bmatrix} \sim \mathcal{N} \left( \mathbf{0}, \\ \begin{bmatrix} \textbf{K}(\mathbf{x}, \mathbf{x}) & \textbf{K}(\mathbf{x}, \mathbf{x}_*) \\ 
    \textbf{K}(\mathbf{x}_*, \mathbf{x}) & \textbf{K}(\mathbf{x}_{*}, \mathbf{x}_{*}) \end{bmatrix} \right).
\end{equation}
Note that \eqref{joint_g} is the joint distribution of the observed pairs ($\mathbf{x}, \mathbf{y}$) and unobserved pairs ($\mathbf{x}_*$, $\mathbf{f}(\mathbf{x}_*)$). The expected value for $\mathbf{f}(\mathbf{x}_*)$ can be derived using conditional properties which leads to
\begin{equation} \label{eq_pred}
    \hat{\mathbf{f}}(\mathbf{x}_*) \triangleq \EX[\mathbf{f}(\mathbf{x}_*) |\mathbf{x}, \mathbf{y}, \mathbf{x}_*] = \textbf{K}(\mathbf{x}_*, \mathbf{x})[\textbf{K}(\mathbf{x}, \mathbf{x})]^{-1}\mathbf{y};
\end{equation}
a complete derivation is given by \cite{rass}. Then, \eqref{eq_pred} can then be used to compute estimates for the value of the function $\mathbf{f}(\mathbf{x}_*)$ at location $\mathbf{x}_*$.

\subsection{Model-Based Clustering}

Clustering, a.k.a.\ unsupervised classification, is an unsupervised machine learning task which attempts to classify (unlabelled) data points into distinct groups. Commonly, clustering is defined as assigning data into groups such that data in the same cluster are more similar to each other than to data in different clusters. Initially this definition seems intuitive; however, practically there are some problems. Namely, grouping each data point into its own cluster would satisfy this definition. Following several others \citep[e.g.,][]{tiedeman55,wolfe63}, \cite{mcnicholas16a} provides a definition not based on similarity: 
``a cluster is a unimodal component within an appropriate finite mixture model'', where the word `appropriate' requires consideration and, specifically, that the component densities have the necessary flexibility to
fit the data \citep[see][for further discussion]{mcnicholas16a}. Whatever definition of a cluster one may prefer, many methods have been developed to tackle this problem of unsupervised learning. Model-based refers to using probability distributions to model the clusters (as opposed to hierarchical clustering, $k$-means clustering, etc.).  

\subsection{Finite Mixture Model}
The finite mixture model is a popular tool for (model-based) clustering --- recent reviews are provided by \cite{bouveyron14} and \cite{mcnicholas16b}. 
The density of a $G$-component finite mixture model is
\begin{align}
f(\mathbf{x} | \bm{\varphi}) &= \sum_{g=1}^G\pi_g f_g(\mathbf{x} | \bm{\theta}_g),
\end{align}
where $\pi_g>0$, with $\sum_{g=1}^G  \pi_g=1$, is the $g$th mixing proportion $f_g(\mathbf{x} | \bm{\theta}_g)$ is the $g$th component density, and $\bm{\theta} = (\bm{\theta}_1, \ldots ,\bm{\theta}_G)$ are the component density-specific parameters, with $\bm{\varphi} = (\bm{\pi}, \bm{\theta})$ and $\bm{\pi}= (\pi_1, ... , \pi_G)$. The likelihood for $\mathbf{x}_1,\ldots,\mathbf{x}_n$ in a model-based clustering paradigm, using a Gaussian mixture model, is given by
\begin{align}\label{mixden}
    \mathcal{L}(\bm{\varphi}) = \prod_{i=1}^n\sum_{g=1}^G \pi_g \ \phi(\mathbf{x}_i \ | \ \bm{\mu}_g, \bm{\Sigma}_g),
\end{align} 
where $\phi(\mathbf{x}_i \ | \ \bm{\mu}_g, \bm{\Sigma}_g)$ is the density of a multivariate Gaussian distribution with mean $\bm{\mu}_g$ and covariance matrix $\bm{\Sigma}_g$.

\subsection{Expectation-Maximization Algorithm}
Model-based clustering requires estimating the unknown model parameters from the likelihood in \eqref{mixden}. The expectation-maximization (EM) algorithm \citep{em1} provides a good starting point for this problem. Each iteration of the EM algorithm  starts by computing the expectation of the complete-data log-likelihood (E-step), then maximizes the conditional expectation of the complete-data log-likelihood (M-step). The E- and M-steps are iterated until some stopping rule is met. Consider a Gaussian model-based clustering complete-data likelihood, denoted by $\mathcal{L}_c(\bm{\varphi})$, where $\bm{\varphi}$ denotes all the parameters and the complete-data comprise the observed $\mathbf{x}_1,\ldots,\mathbf{x}_n$ together with the missing labels $\mathbf{z}_1,\ldots,\mathbf{z}_n$ defined so that $\mathbf{z}_i=(z_{i1}, \ldots,  z_{iG})$ and $z_{ig} = 1$ if observation $i$ belongs to component $g$ and $z_{ig}=0$ otherwise. Now, for the Gaussian mixture model-based clustering paradigm --- corresponding to the likelihood in \eqref{mixden} --- we have complete-data likelihood
\begin{align}
    \mathcal{L}_c(\bm{\varphi}) = \prod_{i=1}^n\sum_{g=1}^G [\pi_g \ \phi(\mathbf{x}_i \ | \ \bm{\mu}_g, \bm{\Sigma}_g )]^{z_{ig}}.
\end{align}
In the E-step, we compute
\begin{align}\label{eqn:zighat}
    \hat{z}_{ig} \colonequals \EX[Z_{ig} |\ \mathbf{x}_i] = \frac{\hat{\pi}_g\phi(\mathbf{x}_i \ | \ \hat{\bm{\mu}}_g, \hat{\bm{\Sigma} }_g)}
    {\sum^G_{h=1}\hat{\pi}_h\phi(\mathbf{x}_i \ | \ \hat{\bm{\mu}}_h, \hat{\bm{\Sigma} }_h)  }
\end{align}
conditional on the current parameter updates (estimates).
Next, in the M-step, the parameters are updated. This amounts to estimating the covariance matrix and mean vector for a Gaussian mixture model. In the M-step, the updates are: 
\begin{align}
 \hat{\pi}_g &= \frac{n_g}{n},\\
 \bm{\hat{\mu}}_g &= \frac{1}{n_g} \sum_{i=1}^n \hat{z}_{ig}\mathbf{x}_i,\\
 \hat{\bm{\Sigma}}_g &= \frac{1}{n_g}\sum_{i=1}^n \hat{z}_{ig}(\mathbf{x_i}-\bm{\hat{\mu}}_g)(\mathbf{x_i}-\bm{\hat{\mu}}_g)^{\top},
\end{align}
where $n_g = \sum^n_{i=1}\hat{z}_{ig}$. After parameter estimation is completed, the clustering results are expressed through the probabilities $\hat{z}_{ig}$, i.e., $\hat{z}_{ig}$ is the probability that $\mathbf{x}_i$ belongs to component $g$. These soft probabilities $\hat{z}_{ig} \in [0,1]$ are often converted into hard classifications via maximum \textit{a~posteriori} probabilities:
\begin{align}
\text{MAP}(\hat{z}_{ig}) =
\begin{cases} 
      1 & \text{if } g = \text{argmax}_h(\hat{z}_{ih}),   \\
      0 & \text{otherwise.}
\end{cases}
\end{align}
Extensive details on model-based clustering and parameter estimation are given by \cite{mcnicholas16a}.

\section{Model to Cluster Functional Data}\label{sec:method}
\subsection{Model Formulation}

We have seen that the log-likelihood for a GP with the observed output vector $\mathbf{y}$ and corresponding input vector $\mathbf{x}$ was distributed according to a multivariate Gaussian distribution. When clustering GPs, the goal will be to find clusters that contain processes which have similar paths. The meaning of similar path refers not only to how close two processes' values are but also to how similar their shapes are (smooth, wiggly, etc.). Now let us define the notation used for the model: the $i${th} GP will have output vector $\mathbf{y}_i$, input vector $\mathbf{x}_i$, and
\begin{align}\label{gp_den}
    p(\mathbf{y}_i | \bm{\theta}_i, \mathbf{x}_i) &= \exp\left\{-\frac{1}{2} \Big(\mathbf{y}_i\textbf{K}^{-1}\mathbf{y}_i^{\top} + \log|\textbf{K}| + n\log2\pi\Big)\right\}.
\end{align}
The density in (\ref{gp_den}) is the probability density function, i.e., $p_g(\mathbf{y}_i | \bm{\theta}_g, \mathbf{x}_i)$, used as the component density for the finite mixture model
\begin{align}
    p(\mathbf{y}_i | \bm{\theta}, \mathbf{x}_i) &= \sum_{g=1}^G\pi_g p_g(\mathbf{y}_i | \bm{\theta}_g, \mathbf{x}_i),
\end{align}
where $\bm{\theta}_g = \{l_g, \sigma_{g}\}$ denotes the hyper-parameters for the $g${th} cluster. 
Because the likelihood of a GP is a Gaussian distribution with covariance matrix $\mathbf{K}$, the complete-data likelihood is given by
\begin{align}\label{likelihood_324}
    \mathcal{L}_c(\bm{\varphi}) = \prod_{i=1}^n\sum^G_{g=1}[\pi_g\phi(\mathbf{y}_i\ |\ \bm{0}, \mathbf{K}_g)]^{z_{ig}},
\end{align}
where $\mathbf{K}_g$ is the covariance matrix corresponding to cluster $g$ and $\phi(\mathbf{y}_i\ |\ \bm{0}, \mathbf{K}_g)$ is the Gaussian density with mean $\bm{0}$ and covariance $\mathbf{K}_g$. An SE covariance kernel is used as the prior on the function space, i.e., 
\begin{align}
    k(x_i, x_j) &=  \sigma^2\exp\left\{-\frac{1}{2l^2} (x_i - x_j)^2\right\}.
\end{align}
The goal is to recover the $G$ pairs of kernel hyper-parameters $\bm{\theta_g} = \{l_g, \sigma_{g}^2\}$ and the mixing parameters $\bm{\pi} = (\pi_1, ..., \pi_G)$, and thence to estimate the latent variables $\bm{z}_1,\ldots,\bm{z}_n$.
 
 \subsection{GP Parameter Estimation}
The first step is to estimate each GP's kernel hyperparameters. 
Herein, the kernel hyper-parameters for the $i$th GP is denoted by $\bm{\Theta}_i=\{l_i$, $\sigma_{i}\}$. In this step, the maximized kernel hyper-parameters for each GP, $l_i^{\text{max}}$ and $\sigma_{i}^{\ \text{max}}$, are estimated. To find these maximized hyper-parameters, an MLE solution is found using gradient ascent, starting with the log-likelihood i.e.,
\begin{align}\label{log_gpden}
    \log p(\mathbf{y}_i | \mathbf{x}_i, \bm{\Theta}_i) &= -\frac{1}{2} \left\{\mathbf{y}_i\textbf{K}^{-1}\mathbf{y}_i^{\top} + \log|\textbf{K}| + n\log2\pi\right\}. 
\end{align}
The derivative is then taken w.r.t.\ to the kernel hyper-parameters
\begin{equation}\begin{split}
    \frac{\partial}{\partial \bm{\Theta}_i}\log p(\mathbf{y}|\mathbf{x}, \bm{\Theta}_i) &=
\frac{1}{2}\mathbf{y}^{\top}\textbf{K}^{-1}\frac{\partial \textbf{K}}{\partial \bm{\Theta}_i}\textbf{K}^{-1}\mathbf{y} -
\frac{1}{2}\text{tr} \left( \textbf{K}^{-1}\frac{\partial \textbf{K}}{\partial \bm{\Theta}_i}\right)
=\frac{1}{2}\text{tr}\left\{\left(\bm{\alpha}\bm{\alpha}^{\top} - \textbf{K}^{-1}\right)\frac{\partial \textbf{K}}{\partial \bm{\Theta}_i}\right\},
\end{split}\end{equation}
where $\bm{\alpha} = \textbf{K}^{-1}\mathbf{y}$. The partial derivatives for $l_i$ and $\sigma_{i}$ are calculated from the first derivatives of the kernel function:
\begin{align}
    \frac{\partial \textbf{K}}{\partial l_i} 
&= \sigma_{i}^2 \exp \Big\{-\frac{1}{2l_i^2} (x_i - x_j)^2 \Big\} (x_i - x_j)^2 l_i^{-3}, \\
\nonumber \\
\frac{\partial \textbf{K}}{\partial \sigma_{i}} 
&= 2\sigma_{i}\exp \Big\{-\frac{1}{2l^2} (x_i - x_j)^2 \Big\}.
\end{align}
After finding the gradient for the likelihood, a gradient ascent algorithm is used to find a sufficiently close solution. This algorithm is given by repeating the following updates until a stopping rule is statisfied:
\begin{align}
l_i^{(t+1)} &= l_i^{(t)} + \lambda \ \frac{\partial}{\partial l_i}\log p(\mathbf{y}_i|\mathbf{x}, l_i^{(t)}),\\ \nonumber
\sigma_{i}^{(t+1)} &= \sigma_{i}^{(t)} + \lambda \ \frac{\partial}{\partial \sigma_{i}}\log p(\mathbf{y}_i|\mathbf{x}, \sigma_{i}^{(t)}),
\end{align}
where superscript $(t)$ denotes iteration $t$.
After maximizing the kernel hyper-parameters, we have 
$\bm\hat{\bTh} = \{\bm\hat{\bTh}_1, \bm\hat{\bTh}_2, \ldots, \bm\hat{\bTh}_n\},$ where $\bm\hat{\bTh}_1=\{l_{1}^{\ \text{max}}, \sigma_{1}^{\ \text{max}}\}$ denotes the maximized kernel hyper-parameters for the first GP, $\bm{\hat{\Theta}}_{2}$ denotes the hyper-parameters for the second GP, and so on.

\subsection{Cluster Parameter Estimation}
The model seeks to cluster the processes and make inferences on the latent variables. A modified EM approach is used. First, the mixing proportions and cluster hyper-parameters are initialized randomly, i.e., we initialize 
$\pmb{\pi}$,
$\pmb{l}$, and
$\pmb{\sigma}$,
where $\pmb{l}=\{l_1, l_2, \ldots, l_G\}$, $\pmb{\sigma} = \{\sigma_{1}, \sigma_{2}, \ldots, \sigma_{G}\}$, and $\pmb{\pi}=\{\pi_1, \pi_2, \ldots, \pi_G\}$ randomly. Next, each GP's ($\text{GP}_1, \dots, \text{GP}_n$) responsibilities are calculated for each of the $G$ clusters to get $n \times G$ responsibilities 
\begin{align}
    \hat{r}_{ig} = \frac{\pi_g \ \phi(\mathbf{y}_i |\bm{0}, \textbf{K}_g)}{\sum^G_{h=1} \ \pi_h \ \phi(\mathbf{y}_i | \bm{0}, \textbf{K}_h)}.
\end{align}
Note that $\hat{r}_{ig}$ represents the responsibility, or conditional expected value, of the $i${th} process belonging to the $g$th cluster, and $\hat{r}_{ig} \triangleq \hat{z}_{ig}$. After the responsibilities are calculated, the mixing proportions $\pi_1,\ldots,\pi_G$ are conditionally maximized on these responsibilities. The update for the $g$th mixing proportion is
$${\pi}_g = \frac{m_g}{m},$$
where $m_g = \sum^n_{i=1} r_{ig}$ is the responsibility for cluster $g$ and $m=\sum_{g=1}^Gm_g$. The cluster-specific kernel hyper-parameters, i.e., $l_g$ and $\sigma_{g}$, are then updated, where $l_g$ is the length-scale parameter for cluster $g$ and $\sigma_{g}$ is the height parameter for cluster $g$:
    $$\hat{l}_g = \frac{1}{m_g}\sum_{i=1}^n \hat{r}_{ig} \ l_i^{\text{ max}},  \qquad\qquad
\hat{\sigma}_{g} = \frac{1}{m_g}\sum_{i=1}^n \hat{r}_{ig} \ \sigma_{i}^{\ \text{max}},$$   
i.e., we are weighting the maximized hyper-parameters, $\sigma_{i}^{\ \text{max}}$ and $l_i^{\text{max}}$, by their respective cluster responsibility $\hat{r}_{ig}$. 

This scheme, for calculating the responsibilities then updating the cluster parameters, is repeated until some stopping rule is met. In this case, when the change in expected complete-data log likelihood at iteration $t$ becomes small, i.e., until
$$\big|\mathcal{Q}(\bm{\varphi}^{(t)}, \bm{\varphi}^{(t-1)}) -\mathcal{Q}(\bm{\varphi}^{(t-1)}, \bm{\varphi}^{(t-2)})\big| < \epsilon,$$
where $\mathcal{Q}(\bm{\varphi}^{(t)}, \bm{\varphi}^{(t-1)})=\EX[\log\left(\mathcal{L}_c(\bm{\varphi}^{(t)})\right)|\mathbf{y}, \bm{\varphi}^{(t-1)}]$ is the expectation of the complete-data log likelihood.

\subsection{Numerical Issues}
At each iteration of gradient ascent, the GP's likelihood gradient needs to be computed:
\begin{align}
    \frac{\partial}{\partial \bm{\Theta}_i}\log p(\mathbf{y}_i|\mathbf{x}, \bm{\Theta}_i)
&=\frac{1}{2}\text{tr}\left\{\left(\bm{\alpha}\bm{\alpha}^{\top} - \textbf{K}^{-1}\right)\frac{\partial \textbf{K}}{\partial \bm{\Theta}_i}\right\}.
\end{align}
This operation requires inverting a $t \times t$ matrix $\mathbf{K}^{-1}$.
Inverting large matrices is notoriously computationally unstable, especially when the matrices are not full rank (or sufficiently close) and eigenvalues become very large or very small. One solution is to first decompose the matrix into lower-triangular form, i.e., 
    $\mathbf{K} =\mathbf{L}\mathbf{L}^{\top}$ and then invert $\mathbf{K}$ via
    $\mathbf{K}^{-1} = (\mathbf{L}^{-1})^{\top}\mathbf{L}^{-1}$.
We use the {\sf R} \citep{R} package {\tt FastGP} \citep{lbb}, which implements the package {\tt RcppEigen} \citep{rcpp} to invert the lower-triangular matrix $\mathbf{L}$. 

\section{Simulation Studies}
This section will first look at two cases of simulated data. The hyper-parameters $\bm{\theta} = \{\bm{l}, \bm{\sigma}\}$ and the mixing proportions $\bm\pi$ will vary based on the simulated sets. The method developed in the previous section will then be applied to recover the hyper-parameters and classify each GP into their respective groups. For the two simulation studies, noiseless squared exponential covariance functions will be used, which in effect means that a perfectly interpolated, noiseless process is observed for the simulation. 

\subsection{Simulation I}

The first simulation starts with generating 30 GPs. The processes are generated on the interval $[0, 10]$ with $T=7$ evenly spaced realizations, i.e., each process has seven values spread evenly on the interval. In all, 10 of the 30 GPs are generated from a multivariate normal distribution using the {\sf R} package {\tt mvtnorm} \citep{mvtnorm}. Where the covariance matrix was constructed using an SE covariance kernel with hyper-parameters $l=1$ and $\sigma = 3$. The remaining 20 were generated similarly but with a covariance matrix constructed with hyper-parameters $l=3$ and $\sigma = 3$.

After running the algorithm described in Section~\ref{sec:method}, estimates for the set of hyper-parameters and mixing proportion were recorded (Table~\ref{tab:table1}). The mixing proportion is easily identified and accurately estimated. Using the MAP classification, the algorithm was able to correctly classify each process. Table~\ref{tab:table1} gives the mean parameter estimates and standard errors. This was done by randomly starting the algorithm 10 times --- i.e., initializing the parameters from a random uniform draw --- and calculating the mean and standard error from these 10 starts.
\begin{table}[ht]
    \caption{Mean values for recovered hyper-parameters, with standard errors, for Simulation~I.}
    \label{tab:table1}
    \centering
\begin{tabular}{lrrr}
\hline
 Parameter & Truth & Mean Estimate & Standard Error\\
\hline
$\pi_1$ & 0.33 & 0.33 & 0 \\  
$\pi_2$ & 0.67 & 0.67 & 0 \\
$l_1$   & 1  & 1.23 & 0.02\\
$l_2$   & 3  & 3.08 & 0.03\\
$\sigma_{1}$   & 3  & 2.18 & 0.07 \\
$\sigma_{ 2}$   & 1  & 1.43 & 0.09 \\
\hline
\end{tabular} 
\end{table}

\begin{figure}[ht]
    \centering
    \includegraphics[scale=.55]{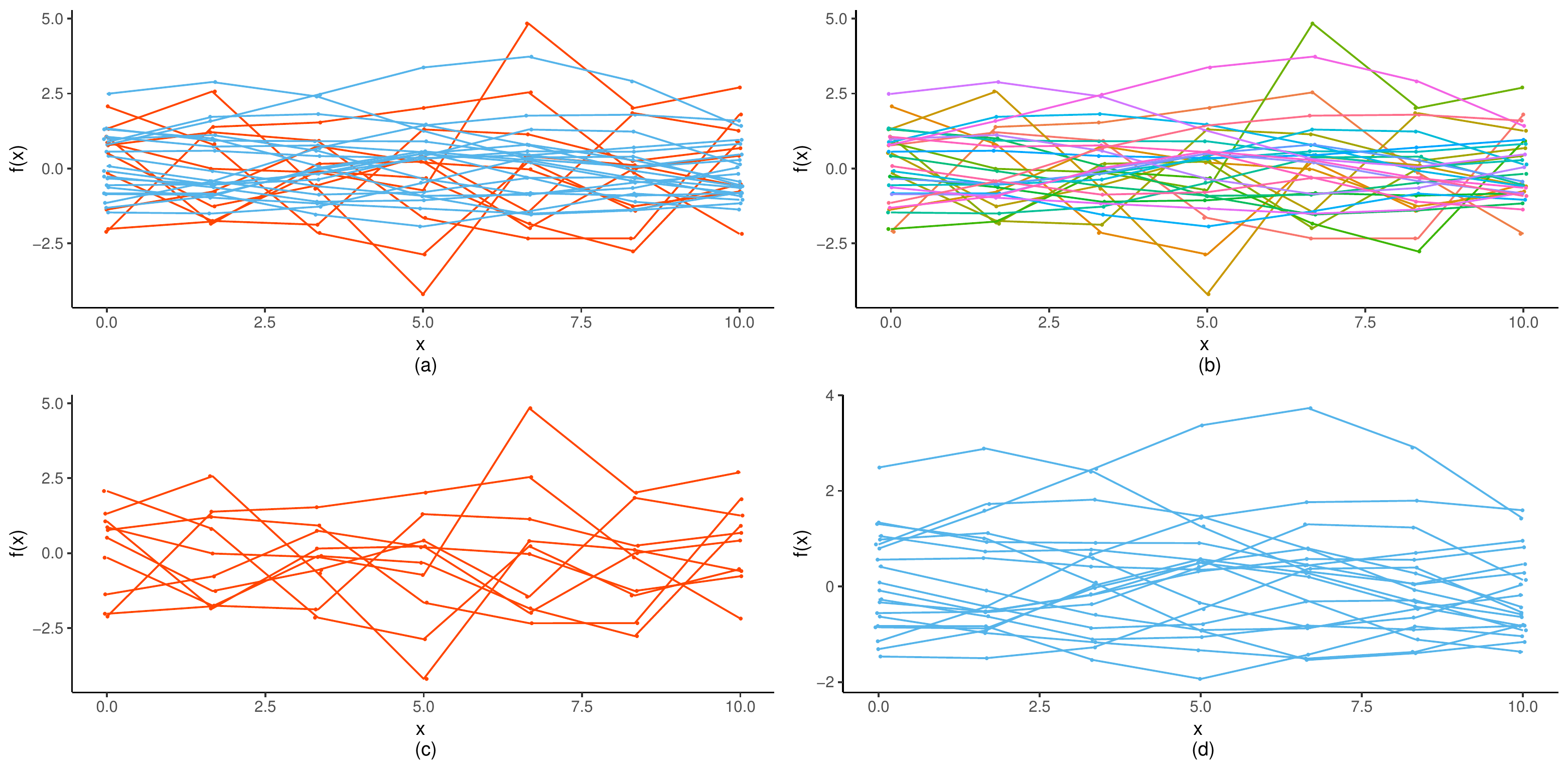}
    \vspace{-0.2in}
    \caption{The 30 GPs from Simulation I: a) coloured by MAP classification, red lines are Cluster~1 and blue lines are Cluster~2. b) Coloured by individual GP. c) GPs from Cluster 1. d) GPs from Cluster 2.}
    \label{fig:gp2}
\end{figure}

Once the processes are coloured by their MAP classification (Figure \ref{fig:gp2}), one can visually see the difference between the two process clusters. The processes ($g=2$, blue) with the larger length-scale $l=3$ are smoother compared to those generated from the process with length-scale $l=1$. 

The length-scale parameter $l$ was also readily recovered in this scenario, producing similar estimates to the true hyper-parameter. The hyper-parameter $\sigma_{1}$, which, recall, controls the function's variance (in $y$, the function's output), is not near the true parameter value. One reason for this could because this cluster has a comparatively small length-scale $l=1$, which models the relative correlation between the points.

\subsection{Simulation II}
\begin{figure}
    \centering
    \includegraphics[scale=.55]{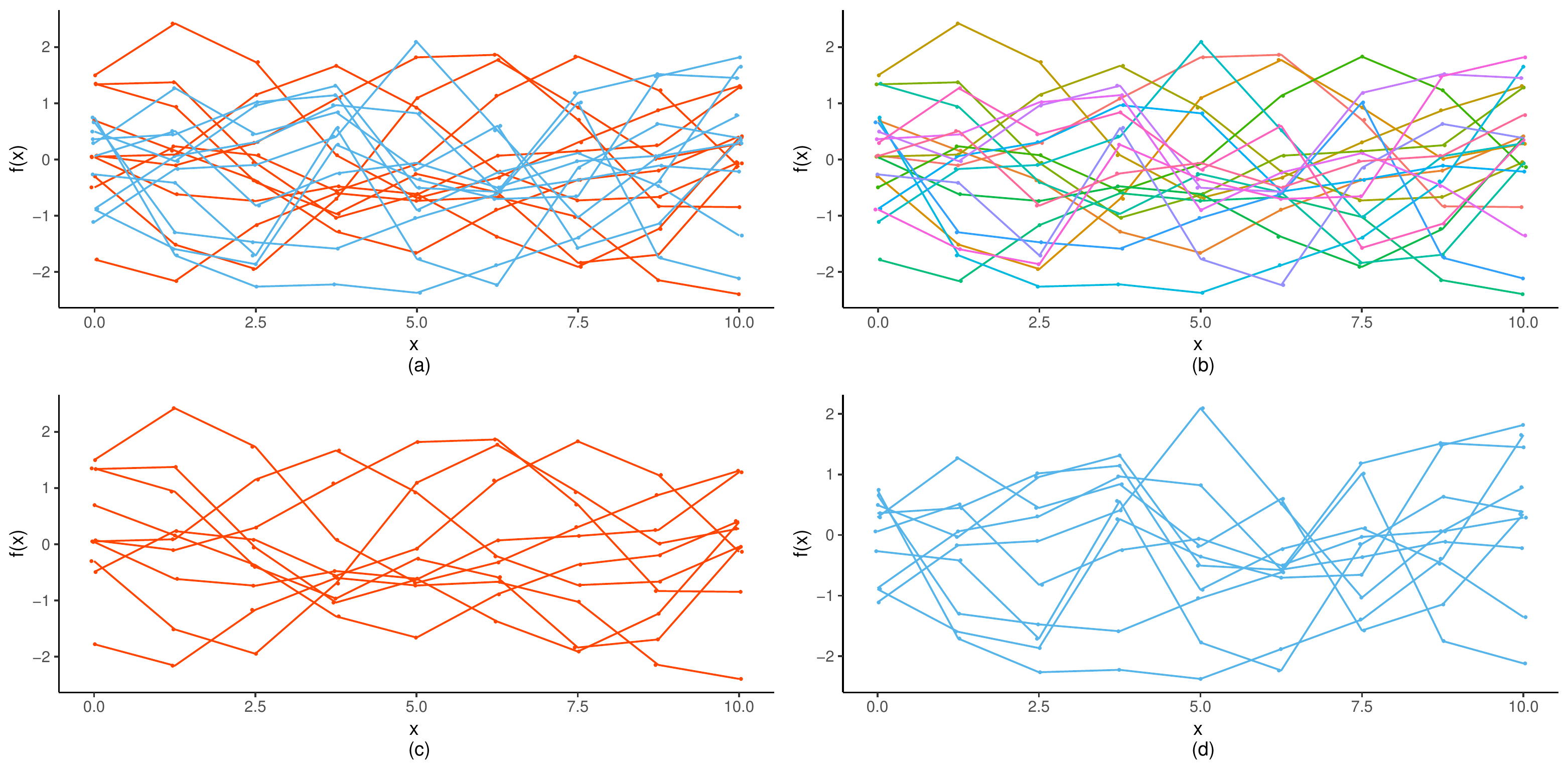}
    \vspace{-0.2in}
    \caption{The 20 GPs from Simulation II: a) Coloured by their MAP classification, red lines are Cluster~1, blue lines are Cluster~2. b) Coloured by individual GP. c) GPs from Cluster 1. d) GPs from Cluster 2.}
    \label{fig:gp_22}
\end{figure}

The second simulation was carried out by first generating 20 GPs. Ten were generated from an SE covariance kernel with hyper-parameters $l=1$ and $\sigma = 1$. The remaining 10 GPs were generated from a covariance kernel with hyper-parameters $l=2$ and $\sigma=2$. Similarly to Simulation~I, the GPs were generated first by constructing the covariance matrix, then by generating random samples using the {\sf R} package {\tt mvtnorm}. In all, $T = 9$ equally spaced observed values were recorded for each GP (Figure~\ref{fig:gp_22}). Based on the plot in Figure~\ref{fig:gp_22}, there seems to be no clear distinction or natural groups of processes. After coloring the processes by their (correct) classifications (Figure \ref{fig:gp_22}), there is still ambiguity about the two groups separation. 

Again, the method accurately recovers the mixing parameter and length-scale (Table~\ref{tab:table_2}). However, for cluster 1, the length-scale $l$ is slightly overestimated and the method inflates $\sigma$ to account for the variance in the function's output. The parameter estimates were calculated by randomly starting the algorithm 10 times and using the mean. Notably, the processes look very similar between groups, so much so that this solution might seem unconvincing if true group labels were unknown.
\begin{table}[ht]
    \centering
    \caption{Mean values for recovered hyper-parameters, with standard errors, for Simulation~II.}
    \label{tab:table_2}
\begin{tabular}{c|c c c}
\hline
 Parameter & Truth & Mean Estimate & Standard Error\\
\hline
$\pi_1$ & 0.5 & 0.52 & 0.004  \\  
$\pi_2$ & 0.5 & 0.48  & 0.004 \\
$l_1$   & 1  & 1.30 & 0.032 \\
$l_2$   & 2  & 2.10 & 0.018 \\
$\sigma_{ 1}$   & 1  & 2.01 & 0.081 \\
$\sigma_{ 2}$   & 2  & 2.05 & 0.082 \\
\hline
\end{tabular}
\end{table}


\section{Coastal Rainfall in British Columbia}\label{sec:bcrain}

\begin{figure}[ht]
    \centering
    \includegraphics[scale=.55]{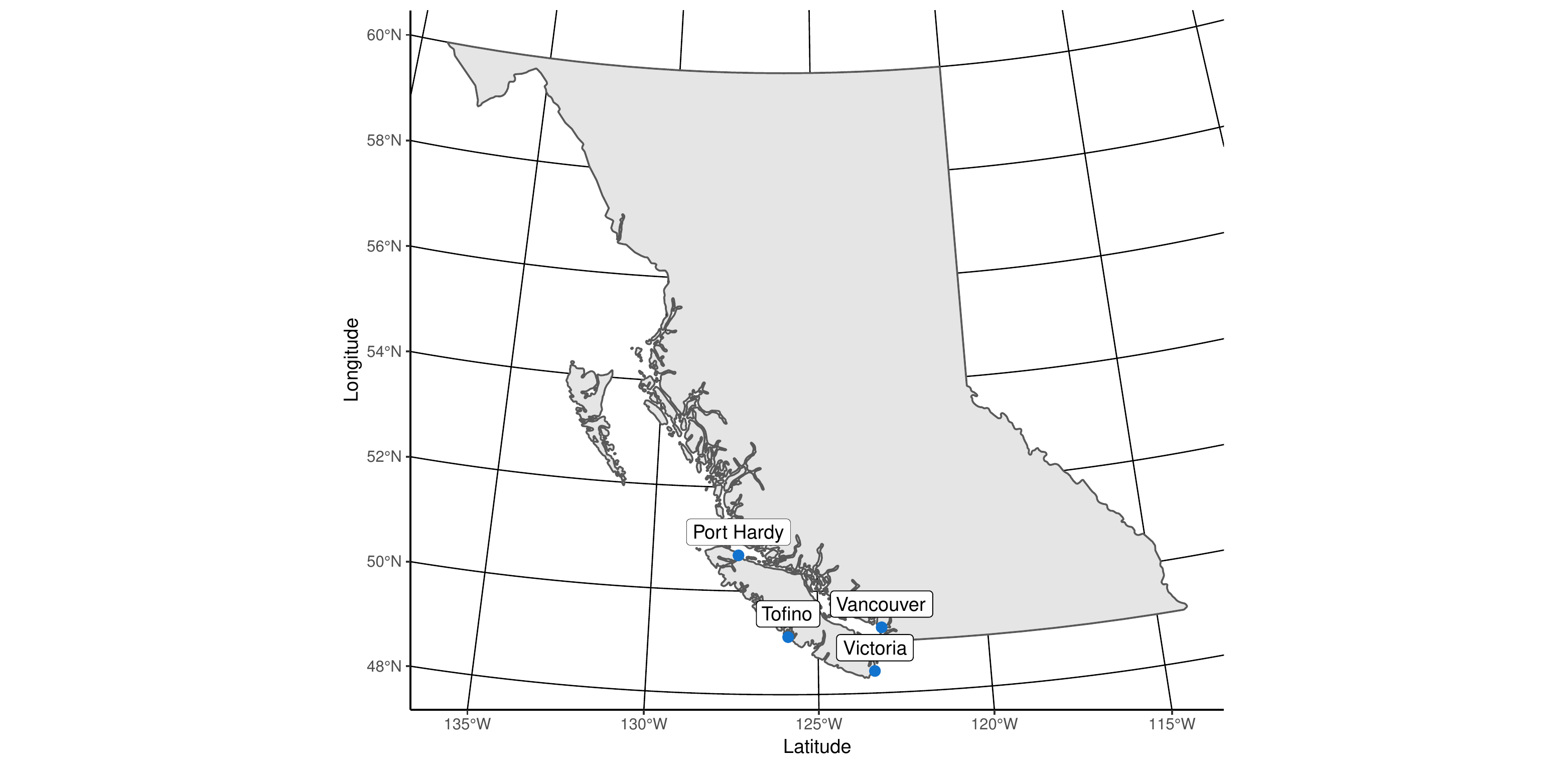}
    \caption{Map of four weather stations, three of which are located on Vancouver Island.}
    \label{fig:tofinotable1}
\end{figure}

This section will look at historical monthly precipitation data for
coastal regions of British Columbia (BC), Canada. These data are recorded by the Government of Canada and collected from the weather stations: Tofino~A, Vancouver International Airport, Port Hardy~A, and Victoria International Airport (Figure~\ref{fig:tofinotable1}).  
These data were derived from the following resources available in the public domain.

\begin{figure}[ht]
    \centering
    \includegraphics[scale=.65]{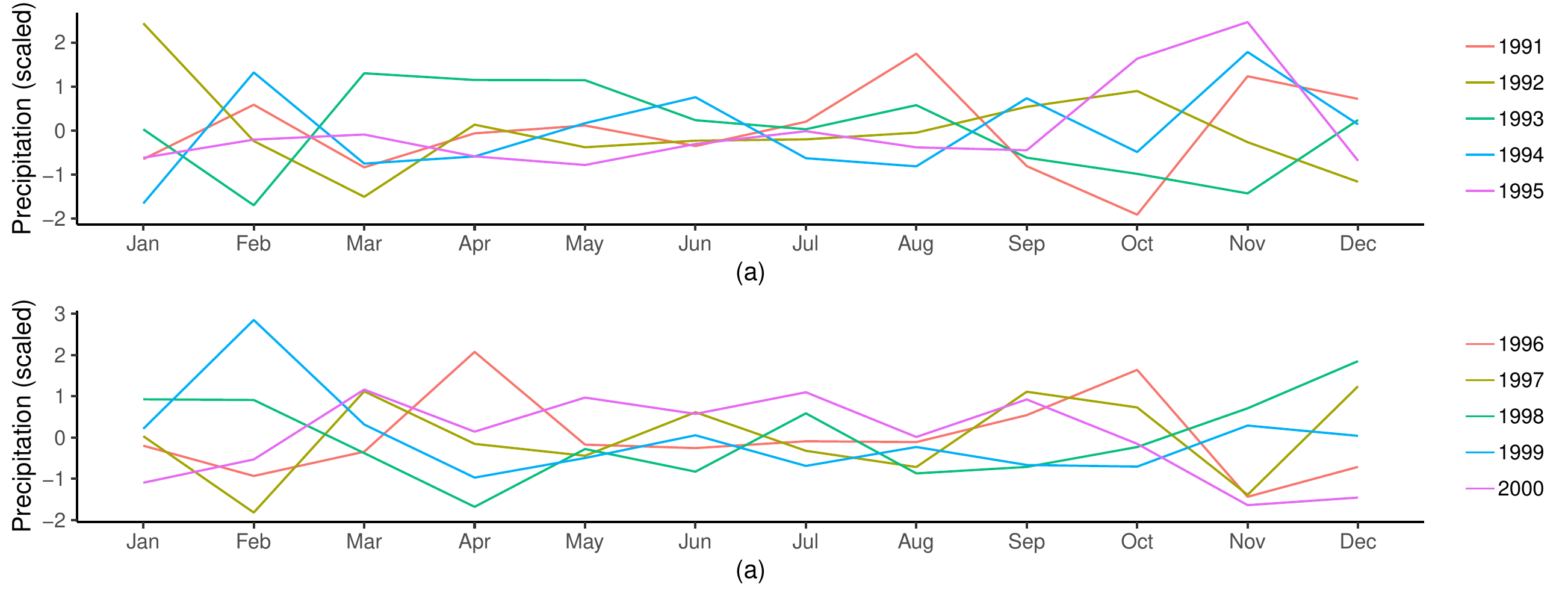}
    \caption{Scaled and season trend removed monthly precipitation for the Tofino coastal region of B.C., Canada. The points are (artificially) connected between months for illustrative purposes, there are 12 measurements per year.}
    \label{fig:tofino_1}
\end{figure}

Total monthly precipitation was recorded from January~1991 to December~2000. The ten years will be treated as independent GPs and each station will be treated separately for clustering purposes. The objective is to find $G=2$ cluster solutions for each station and compare the resulting clusters --- because there are only 10 processes per station, $G>2$ cluster solutions were not explored. The data were processed first by removing the seasonality, i.e., the residual precipitation after a ten year monthly average was removed. The data were then centered and scaled such that the mean is $0$ and standard deviation is $1$. These analyses will use the SE covariance function discussed earlier. Thus, $\bm{\theta} = \{\bm{l}\}$ will be estimated and modelled holding $\bm{\sigma}=1$ constant. Note that the observed outputs $\mathbf{y}$ are (artificially) connected by lines for illustrative purposes (Figure~\ref{fig:tofino_1}). 

\begin{figure}[ht]
    \centering
    \includegraphics[scale=.35]{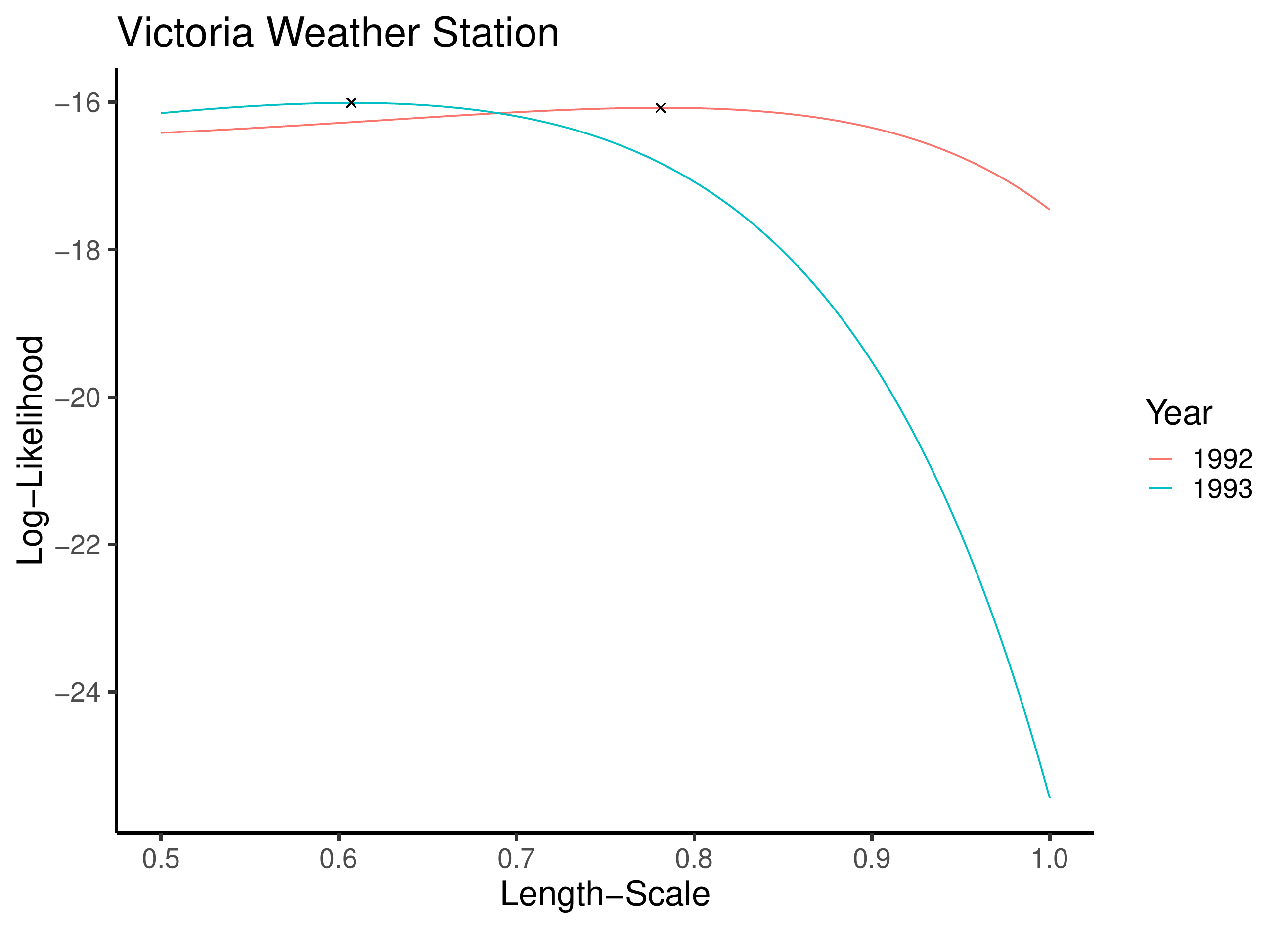}
    \caption{The length scale parameter was chosen by running a grid search over values between 0 and 2. Two years are shown here with each year's respective maximum denoted by the black point.}
    \label{fig:victoria1}
\end{figure}

\begin{figure}[ht]
    \centering
    \includegraphics[scale=.65]{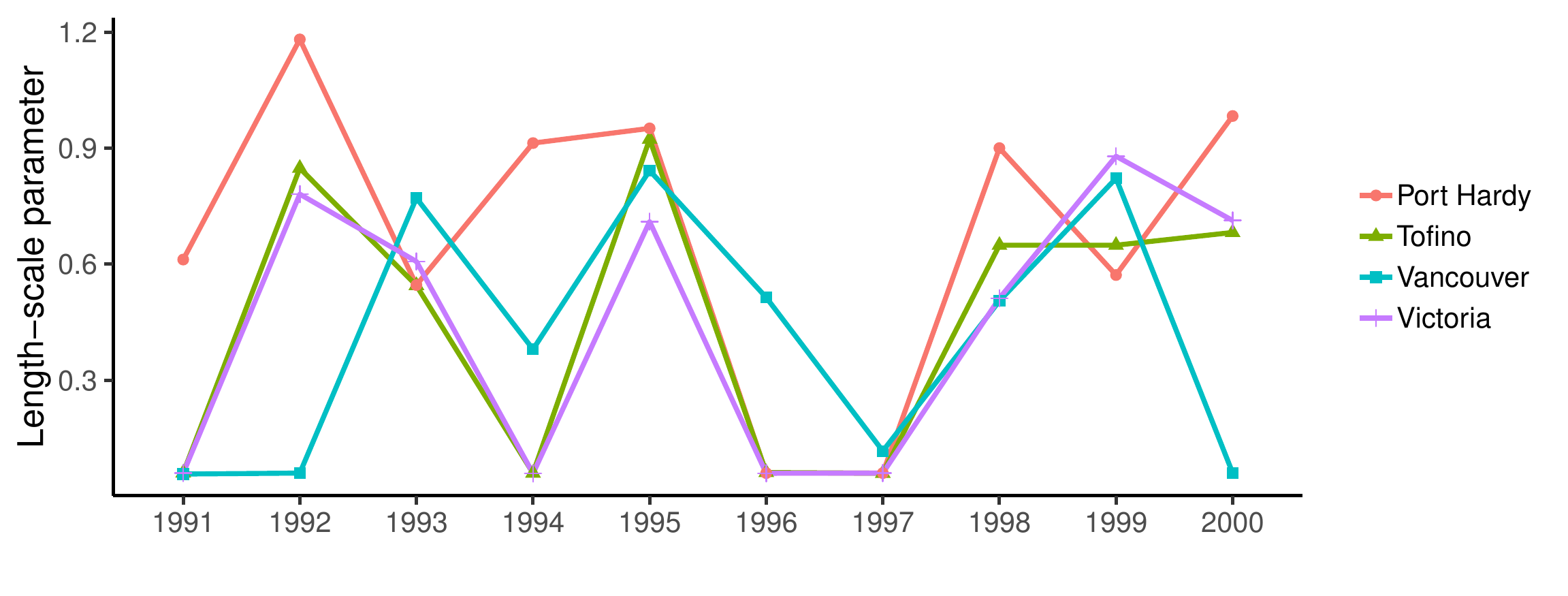}
    \caption{Optimized length-scale hyper-parameter for the ten years of precipitation data, by weather station. The years in Cluster~2 have larger length scale values.}
    \label{fig:pointplot}
\end{figure}

Figure~\ref{fig:tofino_1} illustrates the scaled Tofino data plotted by year. Next, the maximized hyper-parameter $l$ is fitted and results are shown in Figure~\ref{fig:pointplot}. Because of multi-modal likelihoods in the gradient ascent approach to hyper-parameter fitting, a grid search was used to choose the optimal value. An example likelihood from this is shown in  Figure~\ref{fig:victoria1}.

Instead of applying the EM algorithm to estimate mixture parameters, the relatively small number of processes for each station meant that an exhaustive search could be used. That is, each two-group combination of the ten years was considered. This amounted to maximizing $2^{10}=1024$ likelihoods and selecting the model with the highest (complete-data) likelihood.
Using these results, there seem to be two groups emerging, one group with a smaller length-scale and one group with a larger length-scale parameter. Group two has a larger group length scale parameter ($0.842 \leq l_2 \leq 0.985$) as compared with group one ($0.365 \leq l_1 \leq 0.401$). Group two is labelled the ``irregular group" as, with the exception of Port Hardy, the mixing proportion suggests only about $10\%$ of the data come from this group (see Table~\ref{tab:results}). In this case, the years with a smaller length-scale indicate monthly rainfall is less correlated month-to-month than those with a larger length-scale. 
\begin{table}[h!]
  \caption{Years in group two, for each weather station, based on maximizing the complete-data log-likelihood, where remaining years belong to group one.}\label{tab:results_groups}
    \centering
    \begin{tabular}{lr}
\hline
Weather Station & Years in Group Two \\
\hline
Vancouver  &  1995 \\  
Tofino     &  1995 \\ 
Port Hardy &  1992, 1994, 1995, 1998, 2000\\  
Victoria   &  1999 \\  
\hline
\end{tabular}
\end{table}

\begin{figure}
    \centering
    \includegraphics[scale=.55]{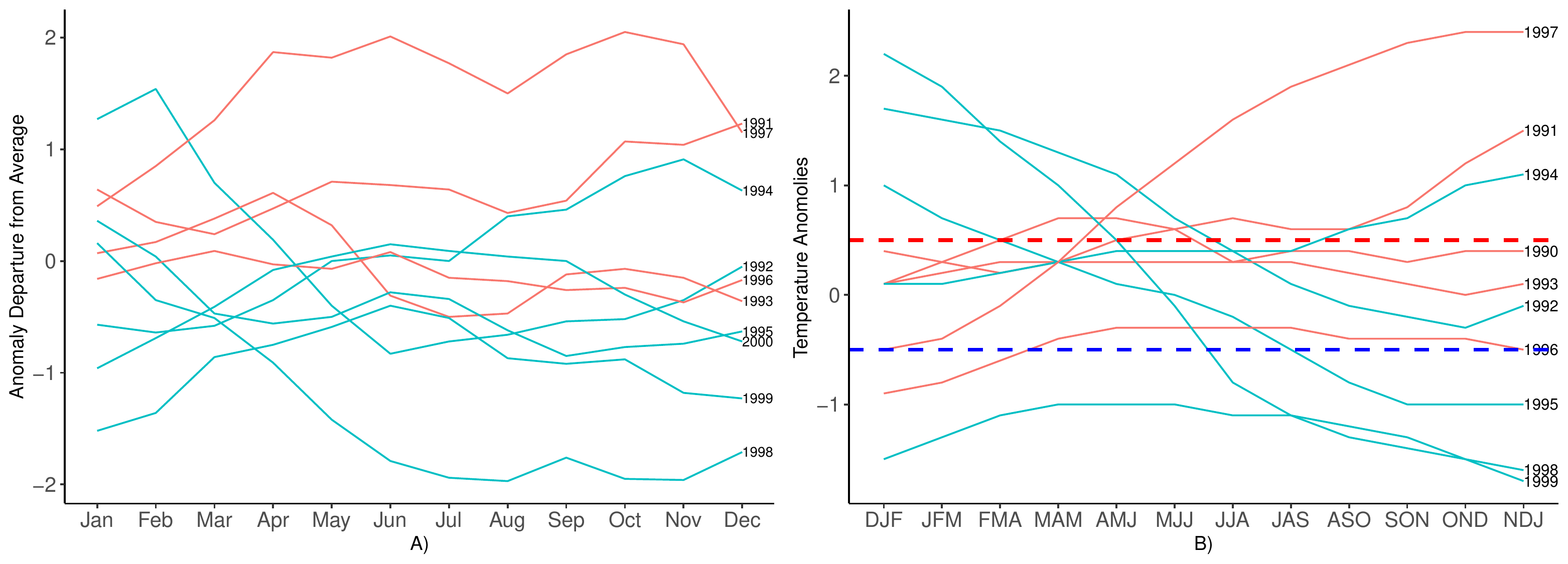}
    \caption{A) Equatorial (160E-80W) upper (surface to 300M) ocean temperature average anomaly based on 1981-2010 climatology. Shows monthly temperature relative to 30 year average. Blue lines (1992, 1994, 1995, 1998, 1999, 2000) represent years that were clustered into group two for at least one weather station. Red lines represent the remaining years. 
Group two years tend to coincide with cooling temperatures. B) Pacific Ocean temperatures 3 month running mean of anomalies for region 3.4 (Middle Pacific Ocean, 5S-5N and 170-120W). Blue lines represent years that at least one weather station clustered into group two (irregular). Red lines represent the remaining years. Warm (red dashed line) and cold (blue dashed line) are a +/- 0.5 threshold for the Oceanic Niño Index (ONI).}
    \label{fig:anoms}
\end{figure}

Vancouver, Tofino, and Port Hardy all had the year 1995 assigned to group two (Table~\ref{tab:results_groups}). Additionally, Port Hardy had the years 1992, 1994, 1998, and 2000 assigned to group two. The years assigned to group two in at least one weather station (1992, 1994, 1995, 1998, 1999, 2000) share an interesting characteristic related to Pacific Ocean temperatures. Specifically, El~Ni\~{n}o and La~Ni\~{n}a are events classified using the Oceanic Ni\~{n}o Index (ONI), an index that measures irregular ocean temperature changes over a three month moving average. An El~Ni\~{n}o (irregularly warm) event immediately preceded a La~Ni\~{n}a (irregularly cold) event twice during the studied time period. Once in 1995 and again in 1998. Both times the year started with warm enough ocean temperatures to classify it as an El~Ni\~{n}o period, and by the end of the calendar year the ocean had cooled enough to be classified as La~Ni\~{n}a \citep{noa}. The other years in the irregular cluster also differed in terms of regular Pacific Ocean temperatures. Generally, years clustered into the irregular group tended to correspond with falling Equatorial Pacific Ocean temperatures  as shown in Figure~\ref{fig:anoms}A). These irregular years also coincided with years where the middle region of the Pacific Ocean (5S-5N and 170-120W) started the calendar year warmer than it ended, as illustrated in Figure~\ref{fig:anoms}B).
\begin{table}[h!]
  \caption{Clustering results, parameters recovered. These estimates are the result of considering each possible two-group combination and selecting the one with the greatest (complete-data) likelihood. The clusters differed mainly with respect to their length-scale parameter $l_1$ and $l_2$.}
\label{tab:results}
    \centering
    \begin{tabular}{ccc}
\textbf{Tofino} \\
\hline
Parameter & Mean Estimate \\
\hline
$\pi_1$   &  0.90 \\  
$\pi_2$   &  0.10 \\ 
$l_1$  &   0.365\\  
$l_2$   &  0.923 \\  
\hline
\end{tabular}
\quad
\begin{tabular}{cc}
\textbf{Victoria} \\
\hline
Parameter&Mean Estimate\\
\hline
$\pi_1$   &  0.90  \\  
$\pi_2$   &  0.10 \\  
$l_1$  &  0.395  \\  
$l_2$   &  0.879\\ 
\hline
\end{tabular} \\[+6pt]
\begin{tabular}{cc}
\textbf{Vancouver} \\
\hline
 Parameter & Mean Estimate \\
\hline
$\pi_1$   &  0.90\\  
$\pi_2$   &  0.10 \\  
$l_1$  &  0.365 \\  
$l_2$   &  0.842 \\ 
\hline
\end{tabular}
\quad
\begin{tabular}{cc}
\textbf{Port Hardy} \\
\hline
Parameter&Mean Estimate\\
\hline
$\pi_1$   &  0.50\\  
$\pi_2$   &  0.50 \\  
$l_1$  &  0.370 \\  
$l_2$   &  0.985 \\ 
\hline
\end{tabular}
\end{table}

From further consideration of the estimated cluster parameters in Table~\ref{tab:results}, Cluster~2's years tend towards a larger length-scale compared to Cluster~1. This suggests that in years where El~Ni\~{n}o changes to La~Ni\~{n}a, rainfall patterns change more smoothly (i.e., are more correlated) across months as opposed to regular weather years.

\section{Discussion}
A method for clustering functional data has been introduced to cluster coastal rainfall data from BC. First, the hyper-parameters that make up a GP were optimized through a gradient-based maximum likelihood optimizer. Because of computational feasibility, parameter estimates were obtained by considering every two-group combination of years and choosing the maximum likelihood. For the simulation studies, the usual EM algorithm for finite Gaussian mixture models was modified. Instead of maximizing the standard covariance matrix, hyper-parameters for a kernel function that measures correlation in $\mathbf{x}$ were optimized. The covariance matrix was then constructed from the optimal kernel parameters. Herein, we fix $G=2$ as known; however, if one were to consider data with more processes per station then $G>2$ could be considered. Notably missing, or incomplete, data can easily be handled by the proposed approach, either by using the predictive distribution to impute the missing data or by ignoring the missing values. This is possible because the model makes inference on the underlying hyper-parameters of the kernel, and not the particular index set of the process.

Two simulation studies were performed. When GPs from different distributions had a large difference in their length-scale parameter $l$ (i.e., 1 versus 3), parameters were readily recovered. When GPs had similar length-scale parameters, $l$ was recovered but $\sigma$ tended to shrink towards a common estimate between both clusters. 
The application to the rainfall data from the coastal region of B.C discovered two groups of years, one which contained ``regular" years and the other ``irregular" years. The irregular years consisted of years where there was a transition from El~Ni\~{n}o to La~Ni\~{n}a, or more generally cooling Pacific Ocean temperatures. These results suggest El~Ni\~{n}o events have some effect on kernel hyper-parameters. The data were standardized to have a zero mean function, implying correlation between rainfall patterns month-to-month can discriminate some El~Ni\~{n}o events (as apposed to magnitude of rainfall). 

The most obvious direction for future work is to apply the approach developed herein to other rainfall data. 
In terms of the BC coastal rainfall data, one could carry out a clustering of all locations combined to see whether some years stand out from others. The BC data could be treated as matrix variate data and clustered accordingly, perhaps after the fashion of \cite{gallaugher18}. In both cases, it is of interest to observe whether $G>2$ clusters emerge. 

\subsubsection*{Acknowledgements}
This work was supported by the Canada Research Chairs program and an E.W.R. Steacie Memorial Fellowship.



\end{document}